\documentclass[sigconf]{acmart}

\usepackage{adjustbox}
\usepackage{multirow}
\usepackage{bm}
\usepackage{algorithm}
\usepackage{algorithmic}
\usepackage{balance}

\copyrightyear{2024}
\acmYear{2024}
\setcopyright{rightsretained}
\acmConference[SIGIR '24]{Proceedings of the 47th International ACM SIGIR Conference on Research and Development in Information Retrieval}{July 14--18, 2024}{Washington, DC, USA}
\acmBooktitle{Proceedings of the 47th International ACM SIGIR Conference on Research and Development in Information Retrieval (SIGIR '24), July 14--18, 2024, Washington, DC, USA}
\acmDOI{10.1145/3626772.3657821}
\acmISBN{979-8-4007-0431-4/24/07}

\makeatletter
\gdef\@copyrightpermission{
 \begin{minipage}{0.3\columnwidth}
 \href{https://creativecommons.org/licenses/by/4.0/}{\includegraphics[width=0.90\textwidth]{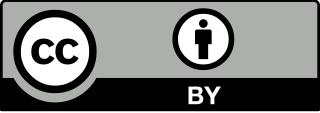}}
 \end{minipage}\hfill
 \begin{minipage}{0.7\columnwidth}
 \href{https://creativecommons.org/licenses/by/4.0/}{This work is licensed under a Creative Commons
Attribution International 4.0 License.}
 \end{minipage}
 \vspace{5pt}
}
\makeatother

\begin{document}


\title{IDGenRec: LLM-RecSys Alignment with Textual ID Learning}


\author{Juntao Tan}
\affiliation{%
  \institution{Rutgers University}
  \city{New Brunswick}
  \country{USA}
}
\email{juntao.tan@rutgers.edu}

\author{Shuyuan Xu}
\affiliation{%
  \institution{Rutgers University}
  \city{New Brunswick}
  \country{USA}
}
\email{shuyuan.xu@rutgers.edu}

\author{Wenyue Hua}
\affiliation{%
  \institution{Rutgers University}
  \city{New Brunswick}
  \country{USA}
}
\email{wenyue.hua@rutgers.edu}

\author{Yingqiang Ge}
\affiliation{%
  \institution{Rutgers University}
  \city{New Brunswick}
  \country{USA}
}
\email{yingqiang.ge@rutgers.edu}

\author{Zelong Li}
\affiliation{%
  \institution{Rutgers University}
  \city{New Brunswick}
  \country{USA}
}
\email{zelong.li@rutgers.edu}

\author{Yongfeng Zhang}
\affiliation{%
  \institution{Rutgers University}
  \city{New Brunswick}
  \country{USA}
}
\email{yongfeng.zhang@rutgers.edu}

\renewcommand{\shortauthors}{Juntao Tan et al.}

\begin{abstract}
    Generative recommendation based on LLMs have transformed the traditional ranking-based recommendation style into a text-to-text generation paradigm. This approach has attracted significant attention. 
    However, in contrast to standard NLP tasks that inherently operate on human vocabulary, current research in generative recommendations struggles to effectively encode recommendation items within the text-to-text framework using concise yet meaningful ID representations. Due to this unresolved issue, the potential of LLM-based generative recommendation systems remains largely unexplored.

    To better align LLMs with recommendation needs, we propose \textbf{IDGenRec}, representing each item as a unique, concise, semantically rich, platform-agnostic textual ID using human language tokens. This is achieved by training a textual ID generator alongside the LLM-based recommender, enabling seamless integration of personalized recommendations into natural language generation.
    Notably, as user history is expressed in natural language and decoupled from the original dataset, our approach suggests the potential for a foundational generative recommendation model. Experiments show that our framework consistently surpasses existing models in sequential recommendation under standard experimental setting. Then, we explore the possibility of training a foundation recommendation model with the proposed method on data collected from $19$ different datasets and tested its recommendation performance on $6$ unseen datasets across different platforms under a completely zero-shot setting. The results show that the zero-shot performance of the pre-trained foundation model is comparable to or even better than some traditional recommendation models based on supervised training, showing the potential of the IDGenRec paradigm serving as the foundation model for generative recommendation. Code and data are open-sourced at \url{https://github.com/agiresearch/IDGenRec}.

\end{abstract}



\maketitle

\section{Introduction}
Generative models with LLMs pre-trained on extensive amounts of information \cite{chowdhery2022palm, chung2022scaling, touvron2023llama, raffel2020exploring} are continually revolutionizing the field of machine learning. Due to their successful in understanding complex instructions and generate creative, contextually relevant predictions, their usage has quickly extended beyond NLP tasks to provide a foundation for applications across diverse research areas. One such example is generative recommendation.

While traditional methods treat recommendation as a retrieval (candidate selection) and ranking process, generative recommendation interprets it as a direct text-to-text generation task: a user's history is expressed as a textual prompt, and the target recommendation is generated in natural language form.
However, unlike NLP tasks that are solely reading and generating human language tokens, items in recommendation platforms are individual entities in an ever-growing universe. Therefore, how to encode items as language tokens (i.e., Item IDs) that can be easily integrated into the text-to-text paradigm is a unique and crucial problem within generative recommendation research.

A few attempts have been made to tackle this problem. P5 \cite{geng2022recommendation}, one of the first works in generative recommendation, proposes allocating out-of-vocabulary (OOV) tokens to items within the recommendation platform. These assigned IDs are fixed-length numerical tokens roughly created based on their sequential appearance in the dataset (e.g., 1001 for the first item, 1002 for the second item). Later, \cite{hua2023index} evaluates how recommendation performance can be further improved by using different strategies to initialize the numerical IDs.

However, while these pioneering works achieve considerable performance in standard recommendation settings, the true capability of LLMs in recommendation remains largely unexplored. First, these methods overlook the wealth of semantic information contained in textual descriptions of items, which undermines one of the primary motivations for using LLMs—harnessing the semantic knowledge gained during their pre-training phase. Second, the numbers assigned to items are meaningless tokens that lack any real contextual meaning. Training such models with a recommendation objective does not lead them to learn the general characteristics of the items. Instead, they merely learn the co-occurrence patterns of these IDs within each dataset. Therefore, although these models present themselves as text-to-text, in essence, they do nothing more than learning representations of each item in a traditional key-value dictionary style, which imposes a ceiling on the quality of the generated recommendations. Simultaneously, since the learned ID representations lack general meanings, the knowledge they acquire is non-transferable across datasets. This means that pre-trained recommendation models lack any zero-shot recommendation ability on unseen data. Consequently, a foundational recommendation model, which has long been pursued in the recommendation community, cannot be achieved by any of the aforementioned methods. 

We propose that the limitations mentioned above primarily arise from inadequate item encoding. Consider tasks in human language, such as question answering and machine translation, where every piece of knowledge is represented within a finite set of tokens. This allows a foundational model to easily learn universally applicable knowledge from training on large text corpus and to adapt effortlessly to any downstream task. If, however, items in recommendation systems were also fully represented using human vocabulary, with each item described by a specific set of natural language tokens, then the capabilities of LLMs could more closely align with the requirements of recommendation systems. In this way, by training on recommendation-specific corpora, LLMs would be able to learn genuine recommendation-related knowledge, which could significantly improve the models' accuracy and generalizability in recommendation tasks.

\begin{figure}[t]
    \centering
    \includegraphics[width=.95\linewidth]{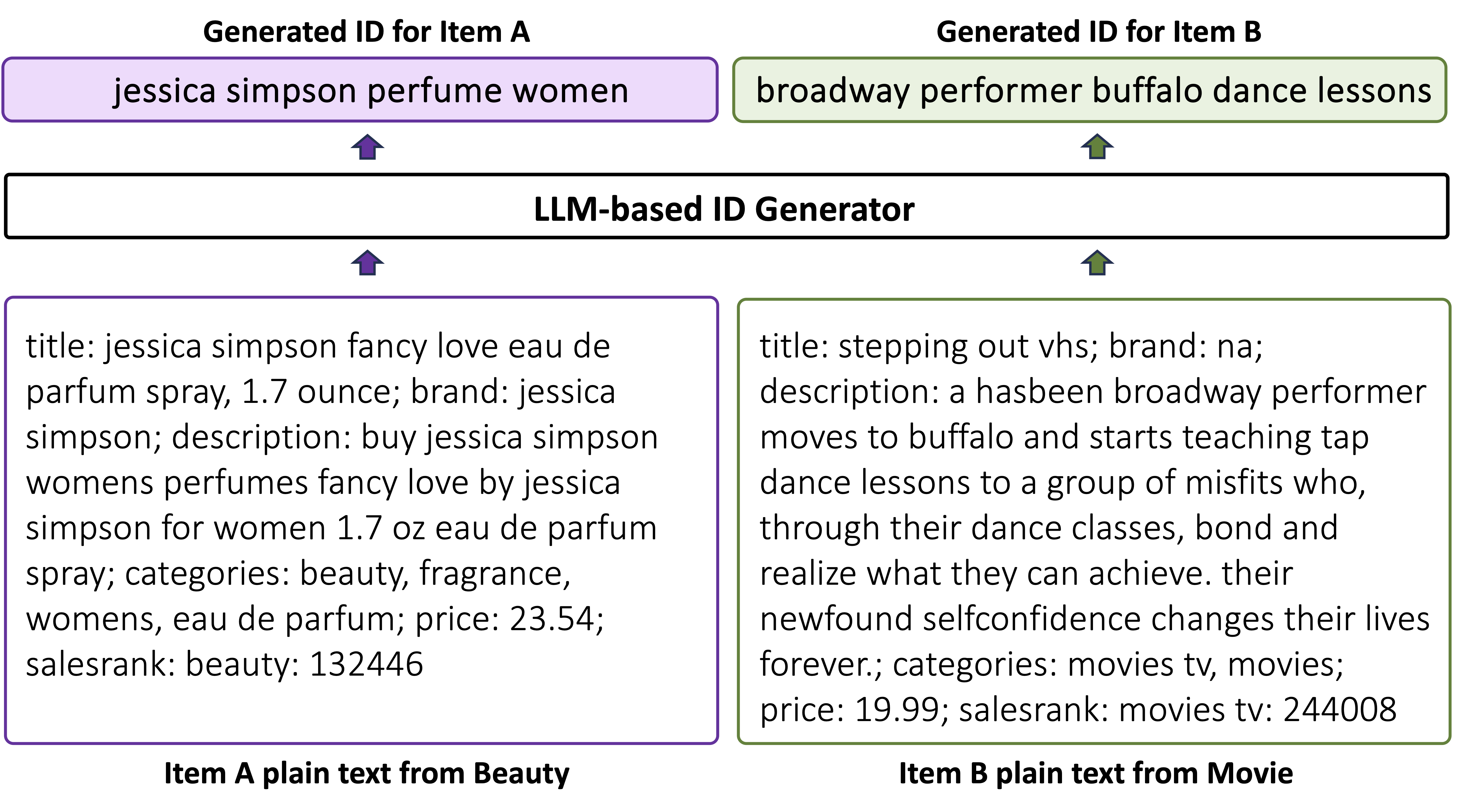}
    \vspace{-10pt}
    \caption{The ID generator takes plain text from each item's meta textual information and generates abstractive textual IDs for the item's representation.}
    \label{fig:id_generator}
\end{figure}

\begin{figure*}[t]
    \centering
    \hspace{-20pt}
    \includegraphics[width=.95\linewidth]{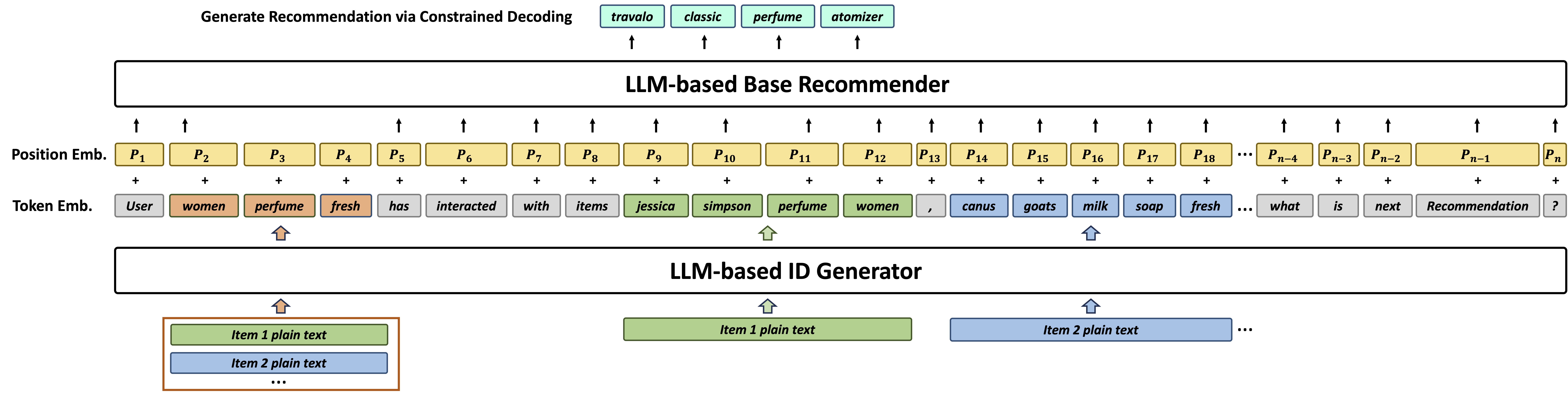}
    \vspace{-10pt}
    \caption{A real example showing the generative recommendation workflow. The ID generator generates item IDs for items from the user history by taking their plain text. Then, the generated IDs are interpolated into the template. Addtionally, the user ID is generated by using all items' text in the user's history, showing a ``high-level profile'' of the user's preference. The position embeddings are subsequently combined with token embeddings to capture the sequence of interactions. Finally, the base recommender generates the ID of the recommended item based on constrained decoding.}
    \label{fig:overview}
\end{figure*}

Hence, we suggest that the ideal IDs in generative recommendation should possess the following properties: 1) They should be textual IDs composed of tokens originally processed by the pre-trained LLMs; 2) They should be meaningful, informative, and suitable for recommendation purposes; 3) The generated IDs should be short yet unique, effectively identifying the recommendation items. However, IDs that meet such stringent requirements are clearly not available in existing item information. Therefore, in this paper, we propose training an ID generator that automatically learns a textual ID for each item that fulfills the above criteria. 
The new framework, named as \textbf{IDGenRec}, treat ID generation as another text-to-text process. As shown in Figure \ref{fig:id_generator}, the ID generator, which is also a language model, takes an item's metadata (i.e., all available textual information about the item) and produces qualified textual IDs. Consequently, the user's history and the target item for recommendation can be represented in natural language, without any ``uncontextualized'' tokens, thus making it suitable for training an LLM-based generative recommender. This overall process is illustrated in Figure \ref{fig:overview}. Notably, by considering all items' text in the user's history, the same ID generator can produce another textual ID, serving as the user's ID that represents a ``high-level profile'' of the user's preferences. The creation of user ID is optional, and we will provide ablation study results in the experiments.

Many challenges lie in this work, and we propose related strategies to address each of them in the paper, including:

\begin{enumerate}
    \item The ID generator should understand lengthy metadata that may include unnecessary information, and should generate tokens that cover the crucial details of the item which are important for recommendations. For this purpose, we have selected a T5 model originally trained for article tag generation and fine-tuned it with recommendation objectives.
    
    \item The generated IDs should be short yet unique, suitable for identifying the recommendation items. However, the automatically generated IDs may not always satisfy the uniqueness criterion, especially as the number of items increases. Therefore, we propose a diverse ID generation algorithm to always ensure each item has a unique ID allocated.
    
    \item Since the framework relies on collaboration between two LLMs---the ID generator and the base recommender---a meticulously designed training strategy is required to enable seamless collaboration between them. We propose an alternate training strategy that trains the LLM-based ID generator and the base recommender asynchronously, ensuring that their learned knowledge is well-aligned.

\end{enumerate}

\begin{table}[t]
\caption{Comparison of LLM-based recommendation models: P5 and its variant versions are generative models but not foundation models due to the use of OOV tokens. UniSRec and Recformer have encoder-only structures and are, therefore, not generative models. Additionally, Recformer employs a rigorously defined item text template that is specifically designed for the Amazon dataset only, and is thus classified as partly a foundation model.}
\vspace{-10pt}
\begin{adjustbox}{width=0.95\linewidth, center}
\begin{tabular}{@{}cccccc@{}}
\toprule
                 & P5 & P5-variants &  UniSRec & Recformer & IDGenRec \\ 
\midrule
Generative Model & \checkmark  & \checkmark         &         &           & \checkmark    \\
Foundation Model &    &                   & \checkmark       & \checkmark (Partly)          & \checkmark    \\
\bottomrule
\end{tabular}
\label{tab:LLM_methods}
\end{adjustbox}
\end{table}

We note that some recent LLM-based recommendation models employ an encoder-only structure, similar to BERT, as their representation function. Some of these models \cite{hou2022towards, li2023text} also possess (or partially possess) characteristics of a foundational recommendation model, though they do not function as generative models. The distinctions among these models are depicted in Table \ref{tab:LLM_methods}. However, generative models present several advantages over discriminative methods. These include transforming the retrieval and ranking processes into a more streamlined generative process, 
eliminating the need of one-by-one item score calculation,
and leveraging the extensive knowledge embedded in pre-trained generative LLMs. Nonetheless, these encoder-only methods remain valuable as baselines for zero-shot evaluation.

We conduct two types of experiments to demonstrate the effectiveness of our proposed method. First, we evaluate the method under a standard sequential recommendation setting. Experiments on $4$ widely-used public datasets show its significant improvements compared to sequential recommendation baselines, including both traditional and generative models. Then, to explore the possibility of training a foundational generative model that learns general recommendation knowledge, inspired by the training paradigm of LLMs in standard NLP tasks, we compile user histories from $19$ datasets from the Amazon Review Datasets, encompassing a diverse range of recommendation domains, to build a massive recommendation training corpus. After training the model on this extensive dataset, we directly apply the foundational model to $6$ unseen datasets either intra- or inter-platform and evaluate the recommendation performance in a completely zero-shot setup. The results show very promising recommendation performance, even surpassing many traditional recommendation models that are based on supervised training.

\section{Approach}
We first introduce the generation process, including how the prompts are constructed and how the generated IDs are integrated into the text-to-text format, as discussed in Section \ref{sec:generative_process}. Then, in Sections \ref{sec:id_generator} and \ref{sec:diver_id_generation}, we describe how the IDs are generated by utilizing the metadata of items, employing a diverse ID generation algorithm to ensure that the IDs are unique for each item. With these generated IDs, the base recommender system is presented in Section \ref{sec:base_recommender}. Finally, in Section \ref{sec:alternate_training}, we demonstrate how the ID generator and recommender are trained alternately with respect to the recommendation objective, ensuring that they work collaboratively and effectively.

\subsection{Generative Process}
\label{sec:generative_process}
In this study, we introduce the foundational generative model within the context of sequential recommendation systems. This approach is particularly apt as it aligns naturally with the sequential representation of user history, serving as an input prompt. Building upon existing work in generative recommendation systems \cite{geng2022recommendation, xu2024openp5, zhang2023recommendation, hua2023index}, our model requires predefined prompt templates for the generation process. For instance, a typical template could be: ``User [user\_ID] has purchased items [item\_ID], [item\_ID], ..., [item\_ID]; predict the next possible item to be bought by the user.'' In this template, all item IDs from the user's history are interpolated at the placeholders, preserving their sequential order. Besides, a user ID can also be produced by the same ID generator, taking the meta information of all items in the user history. Generating a user ID is optional and proved to be benifitial to the recommendation performance in experiments. We have developed $10$ such templates, each with minor differences from the others, randomly selecting one of them for each training instance to ensure the model focuses more on recommendation-relevant information such as the item sequence and less on the exact format of the prompt. An example of a completed prompt is illustrated in Figure \ref{fig:overview}. The decoder then generates the target item ``[target\_item\_ID]'' token by token, which is able to uniquely identify the target recommendation. Later, we will detail the two primary LLM components of our generative process: the ID generator and the base recommender. 

\subsection{ID Generator}
\label{sec:id_generator}
The ID generator is a generative model that produces item IDs using the item's meta-information. This meta-information encompasses all textual data related to the item, including both relevant and irrelevant aspects for recommendation purposes. Potential elements of this information may comprise the item's title, category, price, general description, creation time, popularity, location, etc. The specific content largely depends on the platform and dataset. Although the meta-information is typically presented in a key-value dictionary format, we convert it into plain text during processing, such as ``name: zeppelin; categories: cocktail bars, restaurants; stars: 4.0; ...'' This allows the ID generator to freely learn which pieces of information should be prioritized when generating IDs. 

Consider an item whose plain description is a lengthy sequence of tokens \(\bm{w} = [w_1, w_2, \ldots, w_m]\). Token embeddings will be generated from \(\bm{w}\) by the model's parameters and combined with position embeddings before being fed into the language model. The combination with position embeddings will be omitted in the rest of the formulations since this process is common in large language models. The output of the ID generator will be a concise set of ID tokens \(\bm{d} = [d_1, d_2, \ldots, d_n]\), where \(n \ll m\). When generating each token of the item ID, the model attends to both the item's entire plain description and the previously generated ID units. Thus, the probability of a generated ID is denoted as:
\begin{equation}
    p(d_1,\cdots, d_n) = \prod_{i=1}^n p_\theta(d_i|d_{<i}, \bm{w})
\end{equation}
Where $d_{<i}=[d_1, \cdots, d_{i-1}]$, and $\theta$ are the parameters of the ID generator. This process is illustrated in Figure \ref{fig:id_generator}.

\subsection{Diverse ID Generation}
\label{sec:diver_id_generation}
The two primary properties of the generated IDs are: 1) the IDs should have a reasonable length, and 2) the IDs should be unique. However, these two properties are somewhat contradictory: a set of IDs constrained to a shorter length is more likely to result in duplicates when generated by the ID generator. Therefore, we propose an algorithm to ensure that the generated IDs are both short and unique. The core concept of the algorithm is fundamentally based on diverse beam search (DBS) \cite{vijayakumar2016diverse}, a method commonly used in sentence generation. It is a variation of the standard beam search that is designed to generate a more diverse set of sequences. DBS partitions the beams into groups and introduces a diversity penalty, denoted as $\lambda$, as a hyperparameter to discourage the selection of similar sequences within the same group. A higher value of $\lambda$ promotes more diversity, while a lower value of $\lambda$ gives more weight to the model's probabilities, potentially leading to less diverse outputs.

The ID generator employs DBS in generating IDs. To ensure the uniqueness of the generated IDs, the algorithm generates $k$ groups of IDs each time and compares each generated ID against a set of already existing IDs. If a duplicate is detected, the algorithm increases the diversity penalty in the beam search. This increase in penalty continues until a unique ID is produced, or until the diversity penalty reaches a pre-set maximum threshold (e.g., $10$ in this paper). In cases where the maximum penalty is insufficient to generate a unique ID, the algorithm extends the permissible ID length and repeats this process with the initial diversity penalty. Algorithm \ref{alg:diverse_id_generation} elaborates on this process.

\begin{algorithm}
\caption{Diverse ID Generation Algorithm}
\label{alg:diverse_id_generation}
\begin{algorithmic}[1]
\STATE Initialize set $\mathcal{U}$ to store unique IDs
\STATE Initialize diversity penalty $\lambda$ to $1$
\STATE Initialize ID length limit $L$ to 10
\FOR{each item in the dataset}
    \STATE Initialize $found$ as False
    \WHILE{not $found$}
        \STATE Generate $k$ IDs using ID Generator with current $L$ and $\lambda$
        \FOR{each generated ID $id$}
            \IF{$id$ not in $\mathcal{U}$}
                \STATE Add $id$ to $\mathcal{U}$ and save the item-ID pair
                \STATE Set $found$ to True
                \STATE \textbf{break}
            \ENDIF
        \ENDFOR
        \IF{not $found$}
            \STATE $\lambda \leftarrow \lambda + 1$
            \IF{$\lambda$ exceeds predefined limit}
                \STATE Increase $L$ and reset $\lambda$ to $1$
            \ENDIF
        \ENDIF
    \ENDWHILE
\ENDFOR
\end{algorithmic}
\end{algorithm}

\subsection{Base Recommender}
\label{sec:base_recommender}
After generating item IDs and incorporating them into the prompt template, the system tokenizes the completed prompt and feeds it into the LLM-based recommender. The recommender then generates the tokens of the target recommended item ID in an autoregressive manner. 

To ensure that the decoded ID corresponds to an actual existing item, we adopt a constrained sequence decoding strategy \cite{de2020autoregressive}. More specifically, a prefix tree is used to store all generated candidate IDs. Each newly generated token is constrained by the previously generated tokens, ensuring that the generation process only considers tokens that can potentially form an existing candidate ID in the dataset. Suppose the completed input prompt is denoted as a sequence of tokens $\bm{x} = [x_1, x_2, \cdots, x_n]$. In this case, the base recommender model aims to generate $\bm{y} = [y_1, y_2, \cdots, y_n]$, where $\bm{y}$ is an ID for a real item in the dataset. We define $\mathcal{V}(y_{<i})$ as the subset of valid tokens in the vocabulary, constrained by the prefix tree with previously generated tokens as nodes. The generation of each token by the decoder is defined as:
\begin{equation}
p(y_i|y_{<i}, \bm{x}) =
\begin{cases}
p_\phi(y_i|y_{<i}, \bm{x}) & \text{if } y_i \in \mathcal{V}(y_{<i}), \\
0 & \text{otherwise}.
\end{cases}
\end{equation}

Therefore, the probability of the recommendation of a target item $[y_1, \cdots, y_n]$ is:
\begin{equation}
    p(y_1,\cdots, y_n) = \prod_{i=1}^n p_\phi \big (y_i|y_{<i}, \bm{x}, \mathcal{V}(y_{<i}) \big )
\end{equation}

\subsection{Alternate Training}
\label{sec:alternate_training}

Training the ID generator and the base recommender are two separate but interdependent tasks. The ID generator is trained to produce optimal IDs that the base recommender can easily interpret. Concurrently, the base recommender adjusts its parameters to enhance the correct recommendation for items, each represented by its currently generated ID. Training both components simultaneously may result in an unstable training process. Hence, we propose alternating the training sessions of the ID generator and the base recommender, proceeding through a specified number of iterations. This approach involves asynchronously updating the IDs between the two training phases of the base recommender for better integration and performance.

Both the ID generator and the base recommender are trained by minimizing the negative log-likelihood of the final prediction made by the recommendation pipeline compared to the ground-truth target item ID.

\subsubsection{Training Base Recommender}
At each round of training the base recommender, we pre-compute item IDs for all items using the ID generator at that point. For each user in the training data, the current IDs of the items in the user's history are filled into the sampled template to complete the input prompt $\bm{x}$. Then, the base recommender $\omega$ is trained with a common teacher forcing strategy \cite{williams1989learning}, i.e., the loss of the next token is computed under the ground-truth value of the previous token:
\begin{equation}
    \mathcal{L}_{\text{rec}} = - \sum_{i=1}^{|\bm{y}|} \log P_\omega (y_i|y_{<i}, \bm{x})
\end{equation}

\subsubsection{Training ID Generator}
During this training process, all parameters in the base recommender are fixed, and only the ID generator is updated. The goal is for the ID generator to produce IDs that are suitable for the base recommendation model.

Since the output of the ID generator is a set of discrete tokens (IDs), it is inherently non-differentiable. This poses a challenge for training the model using gradient-based optimization techniques, as gradients cannot flow back through these discrete outputs. To circumvent this, for each item in the user history, we calculate the output logits of each token by the ID generator across all vocabulary, denoted as $\text{Logits}_\phi(\mathcal{V})$, where $\phi$ is the ID generator model and $\mathcal{V}$ is the vocabulary. We then compute the average embedding for each token of the ID through the parameters of the base recommendation model, denoted as $\text{Emb}_\omega\big(\text{Logits}_\phi(\mathcal{V})\big)$, where $\omega$ is the base recommender model. This creates a continuous, differentiable representation of the generated IDs. These ID embeddings are then directly interpolated into the prompt template at the related positions at the embedding level. We use $\text{Emb}_{\text{interp}}$ to represent this completed input embedding. In this way, the ID generator $\phi$ can be trained, guided by the loss computed from the recommendation output. This process is formulated as:
\begin{equation}
\begin{aligned}
    \mathcal{L}_{\text{id}} &= - \sum_{i=1}^{|\bm{y}|}\log P_\omega \left(y_i \mid y_{<i}, \text{Emb}_{\text{interp}}\right) \\
    \text{where } \text{Emb}_{\text{interp}} &= \text{Insert}\left(\text{Emb}_\omega\left(\text{prompt}\right), \text{Emb}_\omega\left(\text{Logits}_\phi(\mathcal{V})\right)\right)
\end{aligned}
\end{equation}
The parameters of the base recommender model (i.e., $\omega$) are fixed, and the loss is only backpropagated to the ID generator (i.e., $\phi$), ensuring that the IDs generated capture the essential characteristics of each item, as determined by their meta-information, in a format that the base recommendation model can effectively interpret.

\subsection{Model Initialization}
We choose the T5 model \cite{raffel2020exploring}
as the backbone for both the ID generator and the base recommender for two main reasons: 1) To maintain the model's simplicity, as this paper does not aim to conduct extensive empirical studies of LLM structures, but rather focuses on the core concept of ID generation; 2) To ensure a fair comparison with previous generative recommendation works, which are also based on T5, thereby demonstrating that the improvement in recommendation ability comes solely from a more elegant ID selection.

For the base recommender, the standard pretrained T5 checkpoint is chosen as the backbone to incorporate pre-learned knowledge into the recommendation task. For the ID generator, given that generating IDs from lengthy texts is non-trivial and highly task-specific, a more dedicated starting point is preferable. Consequently, we select a T5 small model fine-tuned on the article tag generation task\footnote{\url{https://huggingface.co/nandakishormpai/t5-small-machine-articles-tag-generation}}, as the initial configuration for the ID generator. This model, trained on 190k Medium articles\footnote{\url{https://www.kaggle.com/datasets/fabiochiusano/medium-articlesdataset}}, is adept at generating concise tags from article textual content. This selection is driven by the significant similarity between tag generation for news articles and the summarization of items in a few words.

\section{Experiment}
Our experiments comprise two components: the first is an evaluation of standard sequential recommendation to compare the basic supervised learning capabilities of our model against widely-used baselines. The second component is a zero-shot evaluation designed to assess the model's potential as the backbone for a foundational generative recommender system. We will begin by introducing the datasets used in the experiments and detailing our model's training process. Subsequently, we will discuss the experimental results for each of the two experimental settings.

\subsection{Datasets}
For the standard evaluation of sequential recommendation, we selected four widely-used datasets. Three of them, namely Sports, Beauty, and Toys, are from the Amazon review dataset \cite{mcauley2015image, he2016ups}, along with another dataset from Yelp\footnote{\url{https://www.yelp.com/dataset}}. These datasets are also used in previous papers \cite{zhou2020s3, geng2022recommendation, hua2023index}, and we follow the exact data processing steps to filter out users and items with fewer than $5$ interactions, thereby allowing for a fair comparison with all the baselines. 

For the zero-shot experiments for foundational model evaluation, we have two groups of datasets: pre-training datasets and testing datasets. The pre-training datasets are all from the Amazon review dataset, containing various domains, and the testing data are selected from both Amazon review datasets (intra-platform) and the Yelp dataset (inter-platform). We propose a detailed data selection rule for deciding which data are used as pre-training datasets and which are used for testing, as follows:

First, we split all $24$ Amazon review datasets into different groups according to their densities, as shown in Table \ref{table:density_ranges}. Guided by their density range, we include Sports, Beauty, Toys, Music, and Instruments in the test datasets. These datasets cover all the density categories across Amazon review datasets. Besides, including Sports, Beauty, and Toys in the test datasets provides a better view of the foundational model's ability compared to traditional models, since the three datasets are previously used in standard evaluation. Along with Yelp, these form all the test datasets for zero-shot evaluation.

We have $19$ Amazon datasets left, which are used to create the massive recommendation corpus for training the foundational model. Since their sizes vary extremely (e.g., Books contains 603,668 users and Automotive only has 2,928 users), we randomly downsample the large datasets to only include $30,000$ users, which is around the median number of users of the Amazon datasets. All the selected recommendation records together form a ``Fusion'' dataset for training the foundation recommendation model. Complete and detailed data statistics can be seen in Table \ref{tab:dataset}.

\begin{table}[t]
\caption{Amazon review datasets categorized by density}
\vspace{-10pt}
\centering
\begin{adjustbox}{width=0.95\linewidth}
\begin{tabular}{ll}
\toprule
\textbf{Density Range} & \textbf{Amazon Datasets} \\
\hline
$\text{Den.} \geq 0.5$ & Instruments, Patio \\
$0.1 \leq \text{Den.} < 0.5$ & Automotive, Instant, Office, Music, Grocery, Baby \\
$0.05 \leq \text{Den.} < 0.1$ & Tools, Pet, Toys, Phones, Beauty, Games, Apps \\
$\text{Den.} \leq 0.05$ & Clothing, Sports, Health, Home, Kindle, \\
& CDs, Electronics, Movies, Books \\
\bottomrule
\end{tabular}
\end{adjustbox}
\label{table:density_ranges}
\end{table}

\subsection{Implementation Details}

We use SentencePiece \cite{kudo2018sentencepiece} with a vocabulary size of $32,128$ as the tokenizer. The predefined templates for sequential recommendation are largely adopted from P5 \cite{geng2022recommendation}, with the only difference being that the ``dataset'' information is removed from the templates, as our model is more generalized and possesses cross-dataset capabilities.

In the diverse ID generation algorithm, we set $k=10$ as the number of groups for DBS and start with $\lambda=1$ as the initial diversity penalty. This penalty is increased by $1$ in each iteration until the algorithm successfully generates a unique item ID. If $\lambda$ reaches $10$ without producing a valid ID, we then extend the length limit for the item ID. This iterative process of adjusting the diversity penalty and, if necessary, the length of the ID, ensures that the Diverse ID Generation algorithm can successfully produce IDs that are both succinct and distinct. Initially, the token length is set within the range of $[1, 10)$, but if the algorithm is unable to generate a unique ID once the diversity penalty has reached its maximum, we increase the token length limit to the range of $[10,20)$. In our experiments, even for item IDs in the largest ID set, specifically the Fusion dataset, only $11.20\%$ of the items required a second attempt at ID generation with an increased diversity penalty, and merely $3.72\%$ of the items necessitated an extension in ID length.

In the standard recommendation experiments, where the model is trained on a single dataset, we begin by training the ID generator for $1$ epoch, followed by training the base recommender for $10$ epochs, for a total of $3$ iterations. The learning rate is set to $1e-3$ for training the base recommender and $1e-8$ for training the ID generator. This approach is applied to all the datasets in the standard recommendation setting.

Regarding the training of the foundational model for zero-shot recommendation, we find that training the model for one epoch achieved the best performance when tested on the test datasets. Further training negatively affected the zero-shot performance. Interestingly, a similar pattern was also observed in NLP research \cite{touvron2023llama, ouyang2022training}, where pre-training LLMs for more than one epoch led to overfitting on the training data. This observe suggests that the proposed framework narrows the gap between recommendation and language generation.

\subsection{Evaluation Metrics}
In all experiments, we evaluate the ranking performance of the recommendation models using Normalized Discounted Cumulative Gain (NDCG) at 5 and 10, as well as Hit Ratio (HR) at 5 and 10. For fair comparison with baselines, we adopt a leave-one-out strategy for testing. Meanwhile, negative sampling is not used in the evaluation. Instead, we rank over all items for evaluation.
\subsection{Exp1: Standard Evaluation}
\subsubsection{\textbf{Baselines}}
The baselines for standard evaluation include two types of recommendation methods:
\begin{itemize}
\item Traditional sequential recommendation methods, including GRU4Rec \cite{hidasi2015session}, Caser \cite{tang2018personalized}, HGN \cite{ma2019hierarchical}, SASRec \cite{kang2018self}, Bert4Rec \cite{sun2019bert4rec}, FDSA \cite{zhang2019feature}, and S3Rec \cite{zhou2020s3}. These models are popular and cover various model structures, with GRU4Rec based on RNN, Caser on CNN, and SASRec, HGN, Bert4Rec, FDSA, and S3Rec on transformers.
\item Generative recommendation methods, including P5 \cite{geng2022recommendation,xu2024openp5} and its variations \cite{hua2023index} with different ID generation strategies. P5-SID generates numerical item IDs with respect to their sequential appearance order in the dataset.
P5-CID creates item IDs guided by collaborative filtering. P5-SemID uses item category as semantic information to construct item IDs. More details can be seen in \cite{hua2023index}.
\end{itemize}

\begin{table}[t]
\centering
\caption{Dataset Statistics}
\vspace{-10pt}
\label{tab:dataset}
\begin{adjustbox}{width=0.98\linewidth}
\setlength{\tabcolsep}{1pt}
\begin{tabular}{@{}cccccc@{}}
\toprule
\textbf{Category} & \textbf{Datasets} & \textbf{\# Users} & \textbf{\# Items} & \textbf{\# Interactions} & \textbf{Density} \\ 
\midrule
\multirow{4}{*}{Std. Eval.} & Sports & 35,598 & 18,357 & 296,337 & 0.0453\% \\
 & Beauty & 22,363 & 12,101 & 198,502 & 0.0734\% \\
 & Toys & 19,412 & 11,924 & 167,597 & 0.0724\% \\
 & Yelp & 30,431 & 20,033 & 316,354 & 0.0519\% \\ 
\midrule
\multirow{1}{*}{Pre-training} & Fusion & 183,918 & 233,323 & 2,875,446 & 0.0067\% \\
\midrule
\multirow{6}{*}{\shortstack{Zero-shot}} 
 & Sports & 35,598 & 18,357 & 296,337 & 0.0453\% \\
 & Beauty & 22,363 & 12,101 & 198,502 & 0.0734\% \\
 & Toys & 19,412 & 11,924 & 167,597 & 0.0724\% \\
 & Music & 5,541 & 3,568 & 64,706 & 0.3273\% \\
 & Instruments & 1,429 & 900 & 10,261 & 0.7978\% \\ 
 & Yelp (Cross Platform)& 30,431 & 20,033 & 316,354 & 0.0519\% \\
\bottomrule
\end{tabular}
\end{adjustbox}
\end{table}

\subsubsection{\textbf{Results}} The standard evaluation results are shown in Table \ref{tab:standard_evaluation}. Generally speaking, our proposed method significantly outperforms all other baselines. When compared to the second best baseline on each dataset (highlighted with an underline), our method shows improvements of $39.44\%$, $23.55\%$, $42.37\%$, and $36.76\%$ on the Sports, Beauty, Toys, and Yelp datasets, respectively. These improvements are calculated by averaging the increase across four metrics relative to the corresponding best performance from baselines, i.e., the bold value against the underline value of each row for each dataset.
Such results underscore the robustness and significance of the improvements brought by our generative recommendation method. Besides, according to comparisons with P5 and its variants, representing items with learned textual IDs indeed better utilizes the semantic understanding abilities of LLMs.

\begin{table*}[t]
\caption{Standard evaluation for single dataset recommendation. All improvements are significant at $p<0.05$ compared to the best baseline under the student's t-test.}
\label{tab:standard_evaluation}
\vspace{-10pt}
\begin{adjustbox}{width=0.9\linewidth}
\begin{tabular}{@{}ccccccccccccc@{}}
\toprule
Dataset                 & Metric  & GRU4Rec & Caser & HGN & SASRec & Bert4Rec & FDSA & S3Rec & P5-SID & P5-CID  & P5-SemID & \textbf{IDGenRec} \\ \hline
\multirow{4}{*}{Sports} & HR@5    & 0.0129        & 0.0116      & 0.0189    & 0.0233       & 0.0115         & 0.0182     & 0.0251      & 0.0264   & \underline{0.0313}             & 0.0274       & \textbf{0.0429}    \\
                        & NDCG@5  & 0.0086        & 0.0072      & 0.0120    & 0.0154       & 0.0075         & 0.0122     & 0.0161      & 0.0186   & \underline{0.0224}             & 0.0193       & \textbf{0.0326}    \\
                        & HR@10   & 0.0204        & 0.0194      & 0.0313    & 0.0350       & 0.0191         & 0.0288     & 0.0385      & 0.0358   & \underline{0.0431}             & 0.0406       & \textbf{0.0574}    \\
                        & NDCG@10 & 0.0110        & 0.0097      & 0.0159    & 0.0192       & 0.0099         & 0.0156     & 0.0204      & 0.0216   & \underline{0.0262}             & 0.0235       & \textbf{0.0372}    \\ \hline
\multirow{4}{*}{Beauty} & HR@5    & 0.0164        & 0.0205      & 0.0325    & 0.0387       & 0.0203         & 0.0267     & 0.0387      & 0.0430   & \underline{0.0489}             & 0.0433       & \textbf{0.0618}   \\  
                        & NDCG@5  & 0.0099        & 0.0131      & 0.0206    & 0.0249       & 0.0124         & 0.0163     & 0.0244      & 0.0288   & \underline{0.0477}             & 0.0299       & \textbf{0.0486}   \\
                        & HR@10   & 0.0283        & 0.0347      & 0.0512    & 0.0605       & 0.0347         & 0.0407     & 0.0647      & 0.0602   & \underline{0.0680}             & 0.0652       & \textbf{0.0814}   \\
                        & NDCG@10 & 0.0137        & 0.0176      & 0.0266    & 0.0318       & 0.0170         & 0.0208     & 0.0327      & 0.0368   & 0.0357             & \underline{0.0370}       & \textbf{0.0541}   \\ \hline
\multirow{4}{*}{Toys}   & HR@5    & 0.0097        & 0.0166      & 0.0321    & \underline{0.0463}       & 0.0116         & 0.0228     & 0.0443      & 0.0231   & 0.0215             & 0.0247       & \textbf{0.0655}    \\ 
                        & NDCG@5  & 0.0059       & 0.0107      & 0.0221    & \underline{0.0306}       & 0.0071         & 0.0140     & 0.0294     & 0.0159   & 0.0133             & 0.0167       & \textbf{0.0481}    \\
                        & HR@10   & 0.0176        & 0.0270      & 0.0497    & 0.0675      & 0.0203         & 0.0381     & \underline{0.0700}      & 0.0304   & 0.0327             & 0.0376       & \textbf{0.0870}    \\
                        & NDCG@10 & 0.0084      & 0.0141      & 0.0277    & 0.0374       & 0.0099         & 0.0189     & \underline{0.0376}      & 0.0183   & 0.0170             & 0.0209       & \textbf{0.0551}    \\ \hline
\multirow{4}{*}{Yelp}   & HR@5    & 0.0176        & 0.0150      & 0.0186    & 0.0170       & 0.0051         & 0.0158     & 0.0201      & \underline{0.0346}   & 0.0261             & 0.0202       & \textbf{0.0468}    \\ 
                        & NDCG@5  & 0.0110        & 0.0099      & 0.0115    & 0.0110       & 0.0033         & 0.0098     & 0.0123      & \underline{0.0242}   & 0.0171             & 0.0131       & \textbf{0.0368}    \\ 
                        & HR@10   & 0.0285        & 0.0263      & 0.0326    & 0.0284       & 0.0090         & 0.0276     & 0.0341      & \underline{0.0486}   & 0.0428             & 0.0324       & \textbf{0.0578}    \\  
                        & NDCG@10 & 0.0145        & 0.0134      & 0.0159    & 0.0147       & 0.0090         & 0.0136     & 0.0168      & \underline{0.0287}   & 0.0225             & 0.0170       & \textbf{0.0404}    \\ 
\bottomrule
\end{tabular}
\end{adjustbox}
\end{table*}

\subsubsection{\textbf{Ablation Studies}}
We conduct ablation studies regarding two components of the model design.

First, we assess how critical the alternate training strategy is for the model, compared to 1) training only the ID generator while the base recommender remains fixed with the default pre-trained T5 parameters; 2) training only the base recommender with the item IDs directly generated by the T5 model trained on article tag generation. The total number of training epochs is the same as when each is trained in the alternate training setup, i.e., $1 \times 3$ epochs for the ID generator and $10 \times 3$ epochs for the base recommender. The results, shown in Table \ref{tab:ablation_alternate}, indicate that alternate training significantly boosts the model's performance by enabling better collaboration between the two components. Moreover, training only the ID generator does not yield good results. However, training only the base recommender with initially generated IDs still outperforms all baselines.

Second, we evaluate whether generating optional user IDs enhances recommendation performance, as shown in Table \ref{tab:ablation_ID}. In summary, while using only item IDs already allows the model to make excellent recommendations in the standard setting, including user IDs does prove beneficial. Nonetheless, relying solely on user IDs does not provide enough information for making recommendations.

\subsubsection{\textbf{Case Studies}}
\begin{figure*}[t]
    \centering
    \includegraphics[width=0.95\linewidth]{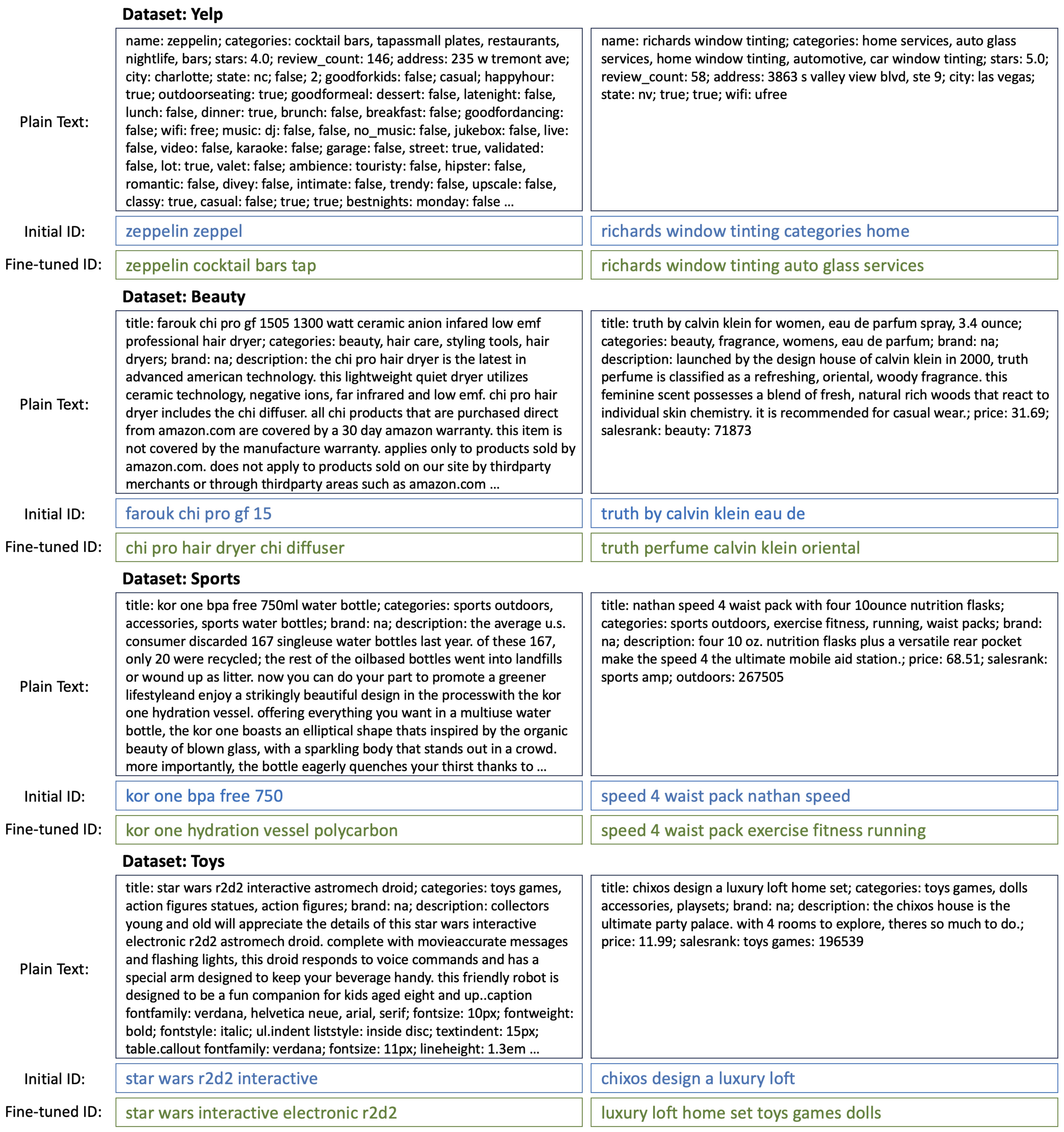}
    \vspace{-10pt}
    \caption{Quality study of the generated item IDs: two examples for each dataset, one with lengthy plain text data and one with shorter plain text data. The blue IDs are generated by the initial ID generator pre-trained on article tag generation, while the green IDs are generated by the ID generator after alternate training.}
    \label{fig:case_ID}
\end{figure*}

As generating item IDs using human vocabulary is the core concept of the proposed method, we conduct an extensive quality study with real examples of the generated item IDs from their plain text information, as shown in Figure \ref{fig:case_ID}. Specifically, we select two examples from each dataset: one with lengthy plain text data and one with shorter plain text data. For each example, we present the ID generated by the initial ID generator and the ID generated by the fine-tuned ID generator at the end of alternate training. Initially, the ID generator tends to select the first few words from the item's meta information, regardless of their relevance. Excitingly, after training, the ID generator is capable of selecting words that more accurately represent the items. The learned IDs are generally more informative and representative, containing less extraneous information such as numbers.

\subsection{Exp2: Zero-shot Evaluation}
We test the potential of our model to serve as a foundation recommendation model. The experiment is conducted by first training the model on the fusion dataset, i.e., an extensive recommendation corpus collected from $19$ recommendation datasets from Amazon Review, then directly test its recommendation performance on each test dataset in a completely zero-shot setting (all the users and items are not seen in the training dataset).

\subsubsection{\textbf{Baselines}}
UniSRec \cite{hou2022towards} is the main comparable baseline in the zero-shot setting, which uses an encoder-only structure to learn item representations from their meta text information. We note that UniSRec incorporates some human-inductive knowledge in data preprocessing, e.g., the ``title,'' ``category,'' and ``brand'' information are carefully selected for Amazon datasets. For the Amazon dataset, we still use the pre-selected categories as in the original paper to showcase its best performance, therefore introducing a slight advantage for UniSRec. For Yelp, we use all the plain text information. Since the patterns have not been seen in both models, this is a fair comparison.

\begin{table}[t]
\caption{Comparison of recommendation accuracy for different training strategies: 1) ID-only: Training only the ID generator. 2) Rec-only: Training only the base recommender. 3) Alternate: Training ID generator and base recommender alternately for $3$ iterations.}
\vspace{-10pt}
\label{tab:ablation_alternate}
\begin{adjustbox}{width=0.8\linewidth}
\begin{tabular}{@{}ccccc@{}}
\toprule
                        &         & ID-only & Rec-only & Alternate  \\ \hline
\multirow{4}{*}{Sports} & HR@5    & 0.0102        &  0.0350        & \textbf{0.0429}       \\
                        & NDCG@5  & 0.0070        &  0.0271        & \textbf{0.0326}       \\
                        & HR@10   & 0.0155        &  0.0461        & \textbf{0.0574}       \\
                        & NDCG@10 & 0.0087        &  0.0307        & \textbf{0.0372}       \\ \hline
\multirow{4}{*}{Beauty} & HR@5    & 0.0111        &  0.0601        &  \textbf{0.0618}           \\
                        & NDCG@5  & 0.0067        &  0.0442        &  \textbf{0.0486}            \\
                        & HR@10   & 0.0192        &  0.0797        &  \textbf{0.0814}         \\
                        & NDCG@10 & 0.0093        &  0.0505       & \textbf{0.0541}          \\ 
\bottomrule
\end{tabular}
\end{adjustbox}
\end{table}

\begin{table}[t]
\caption{Recommendation accuracy for using 1) only user IDs, 2) only item IDs, and 3) both user and item IDs.}
\vspace{-10pt}
\label{tab:ablation_ID}
\begin{adjustbox}{width=0.85\linewidth}
\begin{tabular}{@{}cccccc@{}}
\toprule
                        &         & User ID & Item ID & User \& Item ID   \\ \hline
\multirow{4}{*}{Sports} & HR@5    & 0.0177        & 0.0404         & \textbf{0.0429}                \\
                        & NDCG@5  & 0.0118        & 0.0308         & \textbf{0.0326}                \\
                        & HR@10   & 0.0300        & 0.0528         & \textbf{0.0574}                \\
                        & NDCG@10 & 0.0141        & 0.0348         &  \textbf{0.0372}               \\ \hline
\multirow{4}{*}{Beauty} & HR@5    & 0.0202        & 0.0577         & \textbf{0.0618}                \\
                        & NDCG@5  & 0.0325        & 0.0441         & \textbf{0.0486}                \\
                        & HR@10   & 0.0138        & 0.0778         & \textbf{0.0814}                \\
                        & NDCG@10 & 0.0177        & 0.0506         & \textbf{0.0541}                \\ 
\bottomrule
\end{tabular}
\end{adjustbox}
\vspace{-10pt}
\end{table}

\subsubsection{\textbf{Results}}
The results are presented in Table \ref{tab:zero-shot}. In this zero-shot evaluation, our model generally outperforms UniSRec, with the exception of HR@10 on the Music dataset, where UniSRec demonstrates better performance. Notably, for the Yelp dataset, which is cross-platform from the training data, our model significantly surpasses the baseline with a $353.46\%$ improvement. This demonstrates the superior generalizability of our proposed model, making it more suitable for serving as a foundational model backbone. This is indeed expected, as our method can automatically create representative IDs for the items without requiring any special data processing.

Moreover, by comparing with the average HR and NDCG scores across the shared datasets presented in the standard evaluations, the foundational model's zero-shot recommendation capability is surprisingly comparable to, or even surpasses, that of some traditional recommendation models based on supervised training. For example, the zero-shot performance surpasses that of GRU4Rec on all four datasets, Bert4Rec on three out of four datasets, and Caser on two out of four datasets. Impressively, on the Yelp dataset, our model's zero-shot performance outperforms all traditional methods based on supervised training, only falling short of the P5 model. As the model's zero-shot recommendation capability may further improve with larger and more carefully collected training data, this suggests great potential for its future use as a foundational model.

\begin{table}[t]
\caption{Zero-shot Evaluation. Intra-platform datasets are from the same platform of the training data (i.e., Amazon) and inter-platform datasets are from an unseen recommendation platform (i.e., Yelp).}
\label{tab:zero-shot}
\vspace{-10pt}
\centering
\begin{adjustbox}{width=0.9\linewidth, center}
\begin{tabular}{@{}cccccc@{}}
\hline
Scenario                         & Dataset                      & Metric  &  UniSRec & \textbf{IDGenRec} \\ \hline
\multirow{20}{*}{Intra-platform}   & \multirow{4}{*}{Sports}      & HR@5    & 0.0060   & \textbf{0.0156}       \\
                                 &                              & NDCG@5  & 0.0034   & \textbf{0.0134}       \\
                                 &                              & HR@10   & 0.0098   & \textbf{0.0218}       \\
                                 &                              & NDCG@10 & 0.0046   & \textbf{0.0154}       \\ \cline{2-5} 
                                 & \multirow{4}{*}{Beauty}      & HR@5    & 0.0118   & \textbf{0.0174}       \\
                                 &                              & NDCG@5  & 0.0068   & \textbf{0.0135}       \\
                                 &                              & HR@10   & 0.0206   & \textbf{0.0310}       \\
                                 &                              & NDCG@10 & 0.0096   & \textbf{0.0177}       \\ \cline{2-5} 
                                 & \multirow{4}{*}{Toys}        & HR@5    & 0.0097   &\textbf{0.0103}       \\
                                 &                              & NDCG@5  & 0.0055   & \textbf{0.0079}       \\
                                 &                              & HR@10   & 0.0175   & \textbf{0.0215}       \\
                                 &                              & NDCG@10 & 0.0080   & \textbf{0.0114}       \\ \cline{2-5} 
                                 & \multirow{4}{*}{Music}       & HR@5    & 0.0152   &\textbf{0.0184}       \\
                                 &                              & NDCG@5  & 0.0087   & \textbf{0.0148}       \\
                                 &                              & HR@10   & \textbf{0.0294}   & 0.0238       \\
                                 &                              & NDCG@10 & 0.0133   & \textbf{0.0165}       \\ \cline{2-5} 
                                 & \multirow{4}{*}{Instruments} & HR@5    & 0.0154   & \textbf{0.0203}       \\
                                 &                              & NDCG@5  & 0.0084   & \textbf{0.0139}       \\
                                 &                              & HR@10   & 0.0280   & \textbf{0.0440}       \\
                                 &                              & NDCG@10 & 0.0125   & \textbf{0.0215}       \\ \hline
\multirow{4}{*}{Inter-platform} & \multirow{4}{*}{Yelp}        & HR@5    & 0.0064   & \textbf{0.0300}       \\
                                 &                              & NDCG@5  & 0.0051   & \textbf{0.0248}       \\
                                 &                              & HR@10   & 0.0081   & \textbf{0.0329}       \\
                                 &                              & NDCG@10 & 0.0057   &\textbf{0.0258}       \\ \hline
\end{tabular}
\end{adjustbox}
\vspace{-10pt}
\end{table}

\section{Related Works}
Recent LLM-based recommender systems can be broadly classified into two categories: discriminative and generative \cite{wu2023survey}. 

In LLM-based discriminative recommendation, LLMs are primarily used to learn better representations of users and items by leveraging contextual information in the recommendation process \cite{wu2021empowering, qiu2021u, zhang2022gbert, yao2022reprbert, muhamed2021ctr, xiao2022training, hou2022towards, li2023text, yuan2023go, li2023exploring}. Among them, BERT \cite{devlin2018bert} is commonly used as the backbone. Compared to recommender systems that learn user/item embeddings primarily from user-item associations, these models capture rich textual information in conjunction with collaborative filtering. The core idea of these models is the learning of embeddings with LLMs and integrating the embeddings into 
a ranking score calculation based paradigm. Since this types of works are not the main focus of this paper, readers may refer to \cite{wu2023survey} for a more comprehensive study.

Our work belongs to the second category: LLM-based generative recommendation \cite{geng2022recommendation, geng2023vip5, hua2023index, cui2022m6}. This novel approach transitions from the traditional ranking-based recommendation to a pure text-to-text method. In comparison with the original score computation and re-ranking framework, this method can directly generate the target item from the complete pool of items \cite{li2023large}, thus gathering significant attention. \citeauthor{geng2022recommendation} \cite{geng2022recommendation} introduced one of the first text-to-text generative recommendations, leveraging a pre-trained T5 as the recommendation backbone. The model was fine-tuned on $5$ diverse recommendation tasks with predefined prompts and demonstrated promising results on the downstream tasks. Following \cite{geng2022recommendation}, \cite{geng2023vip5} proposed a multimodal version with a similar architecture, considering item visual features during recommendation generation. In these works, numerical IDs were assigned to each item, enabling the recommended items to be generated token-by-token. Subsequently, \cite{hua2023index} posited that indexing item IDs is the most pivotal aspect of generative recommendation. They conducted an extensive study on indexing methods and assessed their effects on sequential recommendation performance. Although these methods have shown potential, a limitation exists: The rich textual information in recommender platforms is not fully harnessed by using numerical IDs.  Furthermore, the employment of OOV tokens for item representation limits the generalizability of these methods, confining the trained model to a single dataset. Alternative approaches, such as using item titles~\cite{ji2024genrec} or categories~\cite{hua2023index} as semantic IDs, may initially seem to offer a more meaningful representation. However, these approaches may encounter duplicated ID issues as the number of items increases, and the IDs may have unintended overlaps. For example, similar titles may represent completely different items. Our method distinguishes itself by proposing an ID-generator that derives semantic item IDs from rich contextual information, enabling both the utilization of this information and facilitating cross-platform, zero-shot recommendations.

Another branch of generative recommendation research has concentrated on probing the potential of LLMs to directly generate recommendations without the need for training or fine-tuning \cite{liu2023chatgpt, gao2023chat, liu2023first, hou2023large, sun2023chatgpt, dai2023uncovering}. The primary focus in these works is the design of prompts. The emergence of this line of research was inspired by the exceptional zero-shot capabilities of ChatGPT, and they aim to discover how ChatGPT performs in a recommendation scenario. However, as pointed out by \cite{liu2023chatgpt}, ChatGPT cannot generate recommendations with accuracy competitive to that of traditional recommendation methods.

\section{Conclusion}
In this paper, we address the item encoding problem in generative recommendation systems by introducing a novel framework that incorporates textual ID learning. Specifically, we employ an ID generator to produce unique, concise, and semantically rich textual IDs that are platform-agnostic and are based on human vocabulary. We also propose a diverse ID generation algorithm and an alternative training strategy to better align the LLM-based ID generator and the base recommender.
This model has been proven to outperform existing recommendation baselines in the standard sequential recommendation setting. Furthermore, by training our model on some datasets while testing on other unseen datasets, our model shows strong performance under zero-shot recommendation scenario.
Our research offers a new perspective to better align large language models and recommender systems by bridging the two through meticulously learned textual IDs, which may serve as a solid basis for training foundational recommendation models in the future.\\


\noindent\textbf{Acknowledgement}: The work was supported in part by NSF IIS2046457 and IIS-2007907.

\bibliographystyle{ACM-Reference-Format}
\balance
\bibliography{main}

\appendix





\end{document}